# Hiding Information in a Stream Control Transmission Protocol

Wojciech Frączek[1], Wojciech Mazurczyk[1], Krzysztof Szczypiorski[1]
[1]Institute of Telecommunications, Warsaw University of Technology
Warsaw, Poland
Email: wfraczek@gmail.com,{wmazurczyk, ksz}@tele.pw.edu.pl

*Abstract*— The STCP (Stream Control Transmission Protocol) is a candidate for a new transport layer protocol that may replace the TCP (Transmission Control Protocol) and the UDP (User Datagram Protocol) protocols in future IP networks. Currently, the SCTP is implemented in, or can be added to, many popular operating systems (Windows, BSD, Linux, HP-UX or Sun Solaris). This paper identifies and presents all possible "places" where hidden information can be exchanged using an SCTP. The paper focuses mostly on proposing new steganographic methods that can be applied to an SCTP and that can utilise new, characteristic SCTP features, such as multi-homing and multi-streaming. Moreover, for each method, the countermeasure is covered. When used with malicious intent, a method may pose a threat to network security. Knowledge about potential SCTP steganographic methods may be used as a supplement to RFC5062, which describes security attacks in an SCTP protocol. Presented in this paper is a complete analysis of information hiding in an SCTP, and this analysis can be treated as a "guide" when developing steganalysis (detection) tools.

## I. INTRODUCTION

Steganographic techniques have been used for millennia and date back to ancient Greece [4]. The aim of steganographic communication in ancient times and in modern applications is the same: hide secret data (steganogram) in innocent-looking cover material and send it to the proper recipient, who is aware of the information hiding procedure. In an ideal situation, the existence of the hidden communication cannot be detected by third parties. What distinguishes historical steganographic methods from modern ones is, in fact, only the form of the cover (carrier) for secret data. Historical methods used human skin, wax tablets or letters, or other media. Today, steganographic methods use digital media such as pictures, audio, or video, which are transmitted using telecommunication networks. A recent trend in steganography is the utilisation of network protocols as a steganogram carrier by modifying content of the packets, modifying time relations between packets, or using a hybrid solution. All of the information hiding methods that may be used to exchange steganograms in telecommunication networks are described by the term *network steganography*, which was originally introduced by Szczypiorski in 2003 [8]. Many steganographic methods have been proposed and analysed, e.g., [1]-[4]. These methods should be treated as a threat to network security, because they may cause the leakage of confidential information. Steganography as a network threat was marginalised for a few years [20]; however, now not only security staff but also business and consulting firms are becoming continuously aware of the potential dangers and possibilities it creates [10].

Knowledge of the information hiding procedure is helpful to develop countermeasures. Therefore, it is important to identify potential, previously unknown possibilities for covert communication. Such identification is especially important when new network protocols are forecasted to be widely deployed in future networks. For example, detailed analyses of information hiding methods in the IPv6 protocol header were presented by Lucena et al. [9]. In the present paper, we perform similar analyses, except for the use of the Stream Control Transmission Protocol (SCTP) [5]. The SCTP is a transport layer protocol and its main role is similar to two popular protocols, the Transmission Control Protocol (TCP) and the User Datagram Protocol (UDP). The SCTP provides some of the same service features of both, ensuring reliable, in-sequence transport of messages with congestion control. Certain advantages make SCTP a candidate for a transport protocol in future IP networks; the main advantages are that the SCTP is multi-streaming and multi-homing. The popularity of the SCTP is still growing, but it has already been deployed in many important operating systems, such as BSD, Linux (the most popular is *lksctp* [13]), HP-UX, and Sun Solaris. SCTP is supported by the Cisco network device operating system (Cisco IOS) and can even be run in Windows if the proper library is installed [11].

To the best of our knowledge, there are no steganographic methods proposed for the SCTP protocol. However, information hiding methods that have been proposed for the TCP and the UDP protocols (e.g., utilising free/unused or not strictly standard-defined fields) may be utilised in the SCTP as well, due to several similarities between these transport layer protocols and the SCTP. Steganographic methods for TCP and UDP protocols were described by Rowland [1] and by Murdoch and Lewis [2], and very good surveys on hidden communication can be found in Zander et al. [3] and Petitcolas et al. [4].

The main contribution of this paper is to identify and present all possible "places" where hidden information can be exchanged i.e. the whole landscape for the SCTP protocol. This task also includes the identification and presentation of the simplest steganographic methods, e.g., those that substitute the content of certain SCTP header fields, as these methods have been well known for years, given the state of the art. Moreover, even the simplest methods can sometimes be successfully utilised because of ambiguous standardisation, which affects later implementations. For example, padding in Ethernet frames should always be set to zeros, but due to a well-known *Etherleak* [21] effect, more than 20% of Ethernet frames have padding filled with random data [18]. This phenomenon can be utilised to mask hidden communications, even for simple steganographic methods that insert steganograms into padding. In typical cases, such methods will always be easily detectable.

However, the main focus in this paper is on proposing new steganographic methods that utilise new, characteristic SCTP features, such as multi-homing and multi-streaming. When used with malicious intentions, steganographic methods can become perfect tools to launch network attacks. Thus, knowledge about such SCTP-based information hiding solutions can be used as a supplement to RFC5062 [12], which describes security attacks in SCTP protocols and current countermeasures. However, RFC5062 does not include any information about steganography-based attacks and methods of preventing them.

For the vast majority of the presented steganographic methods, modification to the SCTP standard is enough to limit their effectiveness. Proposed in this paper, SCTP-specific steganographic methods can be divided into two groups [18]:

- *Intra-protocol methods*, which may be further divided into the following methods: 1) Modify the content of the SCTP packets, 2) Modify how the SCTP packets are exchanged, and 3) Modify both the content of the SCTP and the way the packets are exchanged, i.e., hybrid methods.
- *Inter-protocol methods,* which utilise relationships between two or more different network protocols to enable secret communication (in our case, the proposed method utilises SCTP and IP protocols).

The above classification is also presented in Fig. 1 and will be used throughout the paper to describe and analyse the proposed SCTP-based steganographic methods. This work is an extension of our previous work [19].

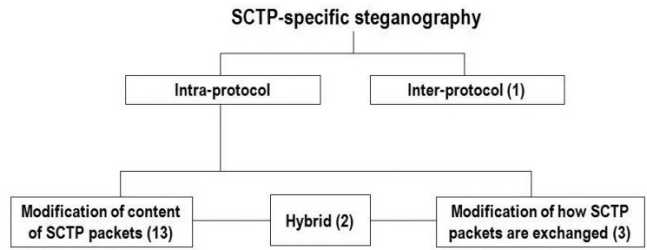

**Figure 1.Classification of SCTP-specific steganographic methods (the number of the proposed steganographic methods for each category is put into brackets)**

The remainder of this paper is arranged as follows. Section 2 gives a brief overview of the SCTP protocol. In Section 3, intra-protocol steganography methods that adopt characteristics of the SCTP protocol are presented. In Section 4, a new inter-protocol method that utilises SCTP is proposed. Section 5 provides possible detection and elimination solutions for the proposed methods. In Section 6, the implementation of one of the proposed methods is described. The methodology of an experiment based on the implemented method is explained in Section 7. Section 8 provides experimental results and analysis. Finally, Section 9 concludes our work.

## II. OVERVIEW OF THE SCTP PROTOCOL

The SCTP [5] was defined by the IETF Signalling Transport (SIGTRAN) working group in 2000 and is maintained by the IETF Transport Area (TSVWG) working group. It was developed for one specific reason - the transportation of telephony signalling over IP-based networks. However, its features make it capable of being a general purpose transport layer protocol ([5], [6]).

SCTP, like TCP, provides reliable in-sequence data transport with congestion control, but it also eliminates the limitations of TCP, which are increasingly onerous in many applications. SCTP also allows users to set order-of-arrival delivery of the data, which means that the data are delivered to the upper layer as soon as they are received (a sequence number is of no significance). Unordered transmission can be set for all messages or for only some of the messages, depending on the application needs.

The SCTP Partial Reliability Extension, defined in [7], is a mechanism that allows users to send only some of the data if all are not necessary, i.e., the data that were not correctly received but became out-of-date. The decision to not transmit some data is made by the sender. He/she has to inform the receiver that some data will not be sent, and the receiver should treat these data as though they had been correctly received and acknowledged. The Partial Reliability Extension and the order-of-arrival delivery enable the use of the SCTP in many applications that are now using UDP.

In TCP, all data are sent as a stream of bytes with no boundaries between messages. This behaviour requires that TCP-based applications have to conduct message framing and must provide a buffer for incomplete messages from the TCP agent. In SCTP, data is sent as separate messages passed by the upper layer. This feature makes SCTP-based applications easier to develop than TCP-based ones.

Each SCTP connection (called association in SCTP) can use one or more streams, which are unidirectional logical channels between SCTP endpoints. Order-of-transmission or order-of-arrival delivery of data are both performed within each stream separately and not globally. If one of the streams is blocked (i.e. a packet is lost and the receiver is waiting for the packet), this blockage does not affect other streams. The benefit of using multiple streams is illustrated in Fig. 2.

As shown in Fig. 2, User X sends four messages (A, B, C, and D) to user Y. There are two requirements concerning the delivery order of these messages. Message A must be delivered before message B, and message C must be delivered before message D. In TCP, messages are sent in the following order: A, B, C, and D, as shown in Fig. 2(1).

If message A is lost, as shown in Fig. 2(2), other messages, in spite of the correct reception, cannot be dispatched to the upper layer until message A is retransmitted and successfully received by user Y, shown in Fig. 2(3). In SCTP, using multi-streaming, messages can be divided into two streams. Messages A and B can be sent within Stream 1, and Messages C and D can be sent within Stream 2, as shown in Fig. 2(4). If message A is lost, as shown in Fig. 2(5), only message B cannot be passed to the upper layer until message A is received. Messages C and D can be delivered to the upper layer, since they are sent within different streams compared to Messages A and B, shown in Fig. 2(6).

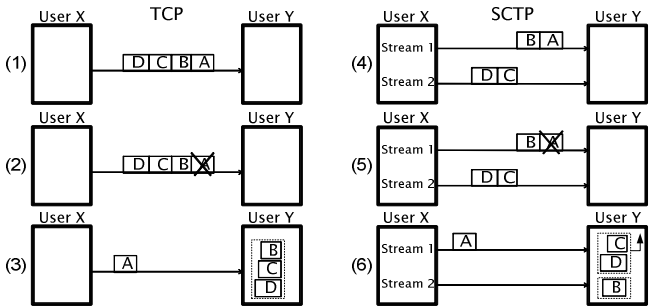

**Figure 2. Comparison of TCP and SCTP data transport using multiple streams**

Another SCTP feature is a provision for protocol extensibility. Each SCTP packet consists of a main header and one or more chunks, as shown in Fig. 3. There are two types of chunks: data chunks that contain user data, and control chunks that are used to control data transfer. Each chunk consists of fields and parameters specific to the chunk type, shown in Fig. 4. Fields are mandatory, and parameters can be either mandatory or optional. SCTP packet structure allows defining not only new chunk types but also broadening functionality of the existing chunk types through defining new parameters.

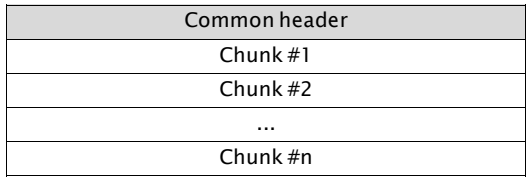

**Figure 3. SCTP packet format**

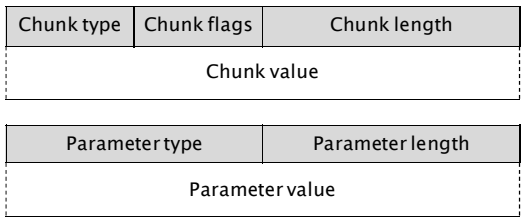

**Figure 4. SCTP chunks and parameters format**

SCTP supports multi-homing, i.e., the host's ability to be visible in the network through more than one IP address, for instance, if the host is equipped with a few NICs (Network Interface Cards). Multi-homing in an SCTP is used to provide increased reliability of data transfer. If there are no packet losses, all messages are transmitted using one source address and one destination address (primary path). If a chunk is retransmitted, the chunk should be sent using a different path (different source and destination addresses) compared to the primary path. Another advantage of SCTP multi-homing in SCTP is the ability to do failover data transfer if the primary path is down.

SCTP uses a four-way handshake with a cookie (Fig. 5), which provides protection against a synchronisation attack (a type of Denial of Service attack); this type of attack is known from TCP implementations. In SCTP, the user initiates an association with the INIT chunk. In response, he/she receives the INIT ACK chunk with a cookie (containing the information that identifies a proposed connection). Then, he/she replies with a COOKIE ECHO, with a copy of the received cookie. Reception of this chunk is acknowledged with the COOKIE ACK chunk. After successful reception of COOKIE ACK, the association is established. Afterwards, connected users can send data using DATA chunks and can acknowledge reception of them with SACK chunks.

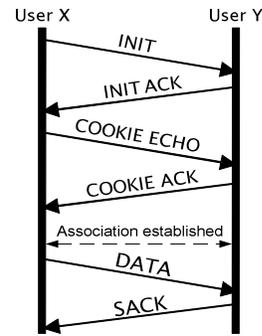

**Figure 5. SCTP association establishment**

Aside from the features described, SCTP also provides a built-in path MTU (Maximum Transmission Unit) discovery, a data fragmentation mechanism; in general, SCTP is considered more secure than TCP.

III. THE PROPOSED INTRA-PROTOCOL METHODS

A. *Methods that modify the content of SCTP packets*

As mentioned above, each SCTP packet consists of chunks, and each chunk can contain variable parameters. We propose 13 new steganographic methods that modify the content of SCTP packets using the following chunks and parameters:
- INIT and INIT ACK chunks – used during initialisation of SCTP association (methods I1, I2),

- DATA chunks – which contain user data (methods D1, D2),
- SACK chunks – used to acknowledge received DATA chunks (methods S1, S2),
- AUTH chunk – used to authenticate chunks (method A1),
- PAD chunk – used to pad packets (method P1),
- Variable parameters – used in specific chunks (methods VP1-5).

These steganographic methods are presented in detail, as follows.

*A.1 INIT and INIT ACK chunks*

(I1) *Initiate Tag* is a 32-bit value of the *Verification Tag* field. This tag must be inserted into each SCTP packet, which is sent to the originator of INIT or INIT ACK chunks within this association. The *Initiate Tag* can be any value except 0 and thus may be used for steganographic purposes, as shown in Fig. 6. The maximum bandwidth of this channel is 32 bits/chunk (fewer bits of this field should be used to limit the chance of detection).

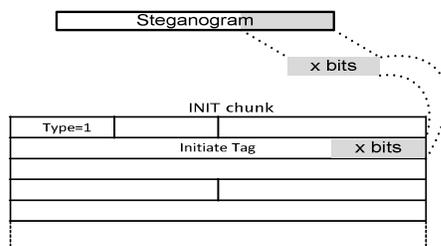

**Figure 6.Method based on the Initiate Tag field**

(I2) *Number of Inbound Streams* is a 16-bit field that defines the maximum number of inbound streams that the sender of the INIT or INIT ACK can handle within this association. In most cases, using more than one hundred streams is unlikely; as a result, at least a few of the most significant bits can be used to insert hidden data. To limit the risk of detection, not only the most significant bits may be used. The potential bandwidth of this method is 8 bits/chunk.

*A.2 DATA chunks*

(D1) *Stream Sequence Number* (SSN) is a 16-bit sequence number within each stream. If the order-of-arrival delivery of the data is set, then there are no requirements concerning the SSN. This feature makes it possible to use the SSN to send steganograms. The maximum bandwidth of this channel is 16 bits/chunk. The method presented can be utilised only if an unordered transmission is set for all data within a stream.

(D2) *Payload Protocol Identifier* is a 32-bit field that represents an upper layer protocol identifier. This field is not used by an SCTP agent because it is for purposes of upper layer protocols. The value 0 indicates no identifier, and other values should be standardised with IANA. SCTP does not verify this value, so it can be used to send secret data. The maximum bandwidth of this channel is 32 bits/chunk.

*A.3 SACK chunks*

(S1) *Advertised Receiver Window Credit* is a 32-bit field that indicates the current size of the SACK sender's receiver buffer. A few least significant bits of this field can be utilised for steganographic purposes. Potential bandwidth of this method is 3-4 bits/chunk; the bandwidth cannot be higher because a higher value could affect flow control.

(S2) *Duplicate TSNs*, which are part of the SACK chunk, are sequence numbers of the duplicate chunks that have been received. This mechanism may enable hidden communication through adding non-duplicating chunk TSNs to the list of duplicate TSNs. In spite of 32 bits of length in the TSN, potential steganographic bandwidth is only a few bits per chunk. This is because adding very different TSNs from recently sent is easy to detect. The method presented here is harder to detect if it is used by multi-homed hosts because it might send SACK chunks with duplicates to addresses other than the source address of the DATA chunks.

*A.4 AUTH chunks*

(A1) *Shared Key Identifier* is a 16-bit field that indicates which pair of shared keys is used in the chunk. This field can be used for covert communication because the receiver of the packet can authenticate the sender through checking all previously exchanged shared keys. The potential steganographic bandwidth of this channel is 1-4 bits/chunk because, in most cases, there will not be many shared keys available. Detection of this method is quite hard because shared keys are established outside of the SCTP protocol.

*A.5 PAD chunks*

(P1) *Padding Data* is a field whose length depends on padding needs. There are no requirements concerning the value of this field, so it can be used for covert communication. Thus, steganographic bandwidth of this channel depends on the size of the padding data.

*A.6 Variable Parameters*

(VP1) IPv4 Address in *IPv4 Address Parameter* and IPv6 Address in *IPv6 Address Parameter* contain addresses of the sending endpoints. These parameters are used for multi-homed hosts and can be attached to INIT, INIT ACK and ASCONF (used to dynamic address reconfiguration) chunks. Each address in these parameters is considered to be unconfirmed until its reachability is not checked. This behaviour allows the use of these parameters for steganographic purposes by sending secret data instead of sending an IP address. However, usage of this method may lead to aborting the association when existing IP addresses are used. The maximum bandwidth is 32 bits/parameter for the IPv4 address and 128 bits/parameter for the IPv6 address.

(VP2) *Heartbeat Info Parameter* is used in the HEARTBEAT chunk, which is exploited to verify reachability of the destination addresses. *Heartbeat Info Parameter* contains the *Sender-Specific Heartbeat Info* field, whose content is not defined, so it can be used as a steganogram carrier. The size of the *Sender-Specific Heartbeat Info* field may differ for various SCTP implementations, which means that the steganographic bandwidth for this method is also implementation dependent. In the Linux Kernel Stream Control Transmission Protocol (lksctp-2.6.28-1.0.10) implementation of SCTP, the *Sender-Specific Heartbeat Info* field has 40 bytes, and thus the steganographic bandwidth for this method is about 320 bits/chunk.

TABLE I. SUMMARY OF VARIOUS METHODS' POTENTIAL STEGANOGRAPHIC BANDWIDTH

| Steganographic method | Steganographic bandwidth | Units |
|---|---|---|
| I1 | 32 | bits/chunk |
| I2 | 8 | bits/chunk |
| D1 | 16 | bits/chunk |
| D2 | 32 | bits/chunk |
| S1 | 3-4 | bits/chunk |
| S2 | 3-4 | bits/chunk |
| A1 | 1-4 | bits/chunk |
| P1 | varies | n/a |
| VP1 | 32 | bits/par. |
| VP2 | 320 | bits/chunk |
| VP3 | 32 | bits/chunk |
| VP4 | 32 | bits/par. |
| VP5 | varies | n/a |

(VP3) *Random Number* in *Random Parameter* can also be used for covert communication. Steganographic bandwidth of this method depends on the purpose of the number. If the number is used in an authentication process, then the random number has 32 bytes and is sent in INIT or INIT ACK chunks. As result, the maximum steganographic bandwidth is 32 bytes/chunk.

(VP4) *ASCONF-Request Correlation ID* in *Add IP Address Parameter*, *Delete IP Address Parameter* and *Set Primary Address Parameter* is a 32-bit field that identifies each request. The only requirement concerning its value is that it be unique for each request; thus, it may be used to transfer steganograms. The maximum steganographic bandwidth of this method is 32 bits/parameter.

(VP5) *Padding Data* in *Padding Parameter* can be exploited for covert communication in the same way as *Padding Data* in *Padding* chunk (see method P1). *Padding Parameter* can be used only in the INIT chunk.

B. *Methods that modify how SCTP packets are exchanged*

*B.1 MULTI-STREAMING*

In SCTP, multi-streaming (for ordered delivery) is realised by utilising two identifiers: Stream Identifier (SI) to uniquely mark the stream, and Stream Sequence Number (SSN) to ensure the correct order of packets at the receiver. Despite these two identifiers, each DATA chunk contains also Transmission Sequence Number (TSN), which is assigned independently to each chunk.

A steganographic method that adopts multi-streaming is based on a determined assignment of TSNs for every chunk distributed along different streams. SIs in subsequent DATA chunks will represent hidden data bits. The example for this method is presented in Fig. 7.

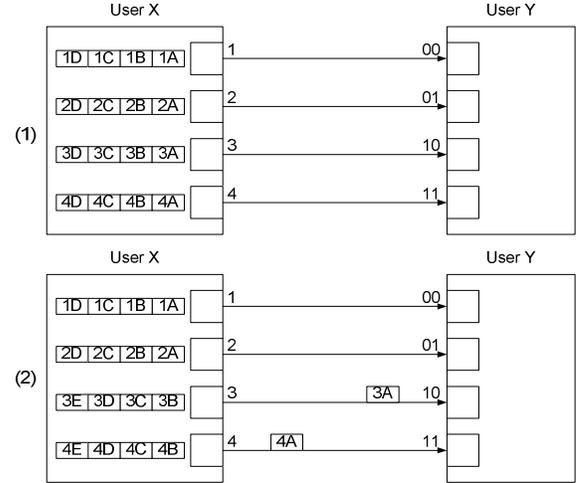

Figure 7. Multi-streaming based steganographic method

At initialisation phase of the SCTP association, users negotiate a number of utilised streams (in the example there are 4 streams). Each stream is assigned with binary sequences shown in Fig. 7(1) – from '00' to '11'. Sending the data through a certain stream depends on the steganogram bits. Therefore, if User X wants to secretly transfer the '1011' bit sequence, then he/she would first send data through Stream 3, then through Stream 4, as shown in Fig. 7(2). It is worth noting that the actual time of the packet arrival is not essential to the correct reception of the secret message. The order of the chunks is determined by the TSN number.

To send steganograms using the proposed steganographic method, it is possible to use not only single chunks but also groups of chunks. Thus, there are $s^k$ ways to send $k$ chunks using $s$ streams. The maximum steganographic bandwidth $S_{MB-MS}$ for this method may be expressed as

$$S_{MB-MS} = \frac{\log_2(s^k)}{k} = \log_2(s) \quad [bits/chunk] \quad (3\text{-}1)$$

According to Eq. (3-1), the maximum steganographic bandwidth is independent of the size of the chunks' group used to hide bits. However, the number of bits actually sent may be different for various sizes of groups. This difference is caused by the fact that the maximum bandwidth may not be a natural number. The greatest influence on the real bandwidth has entier of $\log_2(s^k)$, and the fractional part is less significant. The lower bound of the steganographic bandwidth of this method can be expressed as

$$S_{LBB-MS} = \frac{\lfloor \log_2(s^k) \rfloor}{k} \quad [bits/chunk] \quad (3\text{-}2)$$

If we assume that the number of streams is 3 and that single chunks are used to hide bits of steganogram, one of three values may be assigned to each chunk, for example, "0", "1", "00". Then, the lower bound of the steganographic bandwidth equals 1 bit/chunk ($\lfloor \log_2(3) \rfloor$). This is because it is certain that only "0" or "1" can be sent every time, and the sending of "00" depends on bits of steganogram. If we consider chunk groups of size two, then one of nine values can be assigned to each group; as a result, the lower bound of the steganographic bandwidth equals 1.5 bits/chunk ($\lfloor \log_2(3^2) \rfloor /2$).

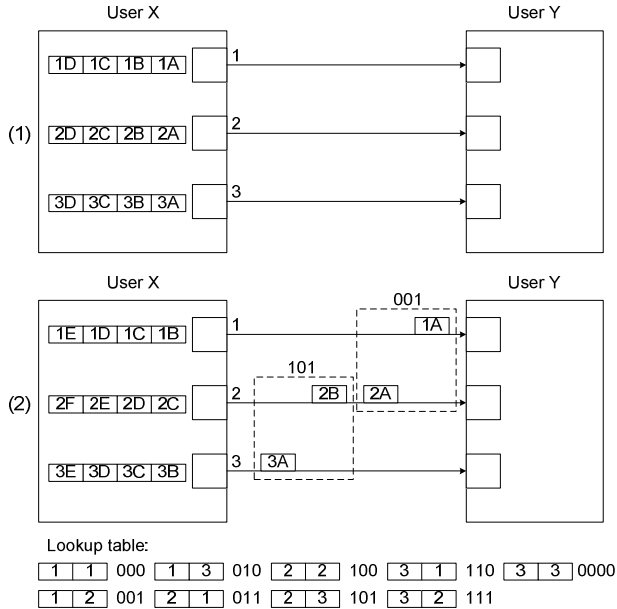

**Figure 8. Multi-streaming based steganographic method – using chunk groups to hide bits**

The example of using chunk groups is presented in Fig. 8. At the initialisation phase of the SCTP association, users negotiate a number of utilised streams (in the example there are 3 streams). In addition, chunk groups are set to a size of 2. At the bottom of Fig. 8, the lookup table, with values assigned to possible chunk groups, is shown. If User X wants to secretly convey the bit sequence "001101", he/she first sends chunks within Streams 1 and 2, then within Streams 2 and 3.

For example, if we assume that the overt communication rate is 250 packets/s, each packet has only one chunk with data and 4 streams are used; then, the steganographic bandwidth is 500 bits/s.

*B.2 CHUNK-BASED FORMAT OF PACKETS*

As mentioned above, each SCTP packet consists of a main header and one or more chunks. This structure allows the sending of a steganogram through the manipulation of messages that are to be transmitted.

The first method, which utilises the modular structure of SCTP packets, modifies the order in which the chunks are to be sent. Each predetermined sequence of chunks may indicate certain steganogram bits. However, this method is limited by two main constraints on the order of chunks in the SCTP packet:

- Control chunks must be placed before DATA chunks.
- DATA chunks must be sent in the order of increasing TSNs.

The result is that only the order of the control chunks may be used for steganographic purposes. The most frequently sent control chunk is SACK, but its broad set of features implies that there is no need to transmit more than one SACK in a single packet. Other control chunks are rarely sent; as a result, the bandwidth of this method is usually less than 1 bit/s. However, if the constraints listed are not considered, then the steganographic bandwidth of this method will increase.

The second method, which uses a chunk-based format for SCTP packets, hides bits in the number of chunks in a single packet. For example, let us assume that a packet consists of a maximum of four chunks. In that case, a packet with one chunk may indicate the bit sequence "00" and a packet with two chunks "01", etc.

The steganographic bandwidth for this method can be expressed as

$$S_{B-FP} = \log_2\left(\left\lfloor \frac{MTU_{data}}{r} \right\rfloor\right) \; [bits/chunk] \qquad (3\text{-}3)$$

where $MTU_{data}$ denotes the value of MTU reduced by the size of the headers of SCTP and the lower layer protocols, and $r$ is the size of a chunk.

For example, if we assume that MTU equals 1400 bytes, and the size of each chunk is 200 bytes, then the bandwidth of this method is about 2.8 bits/chunk. If we additionally assume that the overt communication rate is 250 packets/s, then the steganographic bandwidth is about 700 bits/s.

*C. Hybrid methods*

The SCTP partial reliability extension was also proposed by Stewart et al. [7]. This extension allows the system to *not* retransmit certain data despite the fact that the data was not successfully received. This construct is possible through the FORWARD TSN (FT) chunk, where a new acknowledgement of TSN is inserted. After receiving such a message, the receiving side treats the missing chunks with equal or lower TSNs, as they were properly delivered. This functionality may be adopted for steganographic purposes.

The idea of the first proposed method is similar to the concept of LACK (Lost Audio Packets Steganography), which was developed for real-time multimedia services by Mazurczyk and Szczypiorski [14].

The main idea of this method is presented in Fig. 9. From the User X data, the chunk sent with TSN 6 is skipped, and into this chunk the steganogram is inserted (1). Next, User X sends an FT chunk to signal a new acknowledgment of the TSN (2). After successful reception of an FT chunk, User Y issues a SACK chunk with a new acknowledgment of the TSN (3). When User X receives the SACK chunk, he/she can send an omitted DATA chunk with a steganogram (4). Presented behaviour may look suspicious for many applications, which is why this steganographic method should be used in applications that frequently send large messages.

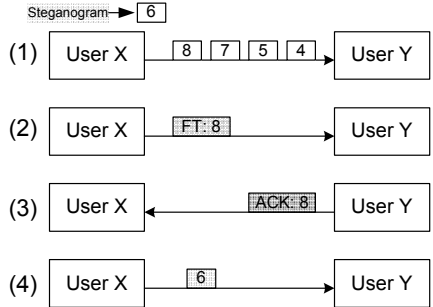

Figure 9. First hybrid SCTP steganographic method

If we assume that the covert communication rate is 250 packets/s, each packet has only one chunk, with a payload size of 1000 bytes. If we use 0.01% of the packets to insert a steganogram, then the potential steganographic bandwidth is 200 bits/s.

The idea of the second method is presented in Fig. 10.

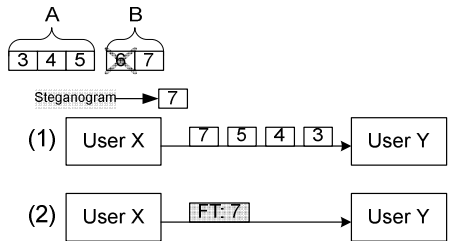

Figure 10. Second hybrid SCTP steganographic method

User X is to send messages A and B. Both messages are too large to put into a single packet, so they are fragmented (message A into chunks with TSNs of 3, 4, and 5, and message B into chunks with TSNs of 6, and 7). Message B should not contain user data and can be inserted intentionally to send the steganogram. User X puts the steganogram into chunks with TSN 7 and sends the chunks with TSN 3, 4, 5, and 7 (without the chunk with TSN 6) to User Y. Then, User X sends the FT chunks to signal a new acknowledgment to the TSN. After User Y receives the FT chunk, he/she analyses chunks in the buffer and finds part of an incomplete message (chunk with TSN 7), which contains a steganogram.

If we assume that the covert communication rate is 250 packets/s, then the size of the messages generated by the user is 2800 bytes (forcing fragmentation into two chunks of 1400 bytes each). We use 0.01% of the packets to insert the steganogram. As a result, the potential steganographic bandwidth is 280 bits/s.

## IV. THE PROPOSED INTER-PROTOCOL SCTP-BASED METHOD

The SCTP multi-homing feature can be utilised to perform hidden communications. The main idea of the proposed steganographic method is presented in Fig. 11. Two users establish SCTP association (User 1 and User 2), and each of them is equipped with more than one NIC. The primary path for the users' communication is through interfaces A and X (1). If $n_1$ denotes the number of the alternative senders' NIC addresses (in Fig. 11 there are 2), and $n_2$ represents the number of alternative receiver NIC addresses (in Fig. 11 also 2), then each address can be used to represent one steganogram bit (or a sequence of bits). Possible alternative paths for communication between these users are as follows: BY, BZ, CY and CZ. User 1's B interface IP address represents binary '0', and the C interface IP address represents binary '1' (a similar situation occurs for User 2). Assigning the bits or the sequences of bits to the users' NICs may depend on the IP address values, i.e., the available NIC addresses can be sorted from the lowest to the highest, and then consecutive values (bit sequences) can be assigned.

If User 1 wants to send a steganogram, then he/she waits for the transmission error on the primary path to occur and then retransmits a chunk through the appropriate path. For example, in Fig. 11, if User 1 wants to send a steganogram that consists of the sequence '01', he/she waits for the transmission error on the primary path to occur (1) and sends retransmitted packets through path BZ (2). Before sending a steganogram, it should first be established which retransmitted chunks carry hidden data. Users can assume that all retransmissions carry bits of steganogram or should mark the beginning of a hidden communication, for example, with an initiation sequence (a sequence of retransmitted chunks through paths that were agreed on previously).

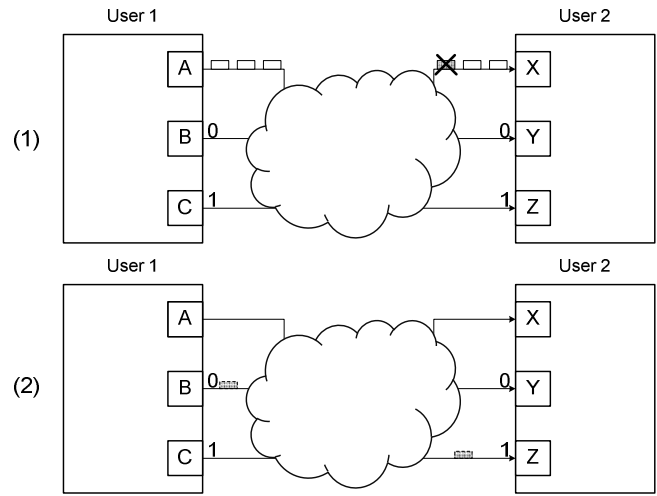

Figure 11. Multi-homing based steganographic method

If retransmitted chunks (retransmission is a SCTP feature) are sent using alternative paths (multi-homing mechanisms available in SCTP), then proper IP addresses must be set in an IP packet. The above-mentioned features of the presented method cause the property that the method's functioning is not limited to only the SCTP protocol. Instead, this method is an example of inter-protocol steganography that utilises relationships between SCTP and IP protocols.

Steganographic bandwidth $S_{B-MH}$ for this method can be expressed as

$$S_{B-MH} = log_2(n_1) + log_2(n_2) \; [bits/chunk] \quad (4-1)$$

For example, in Fig. 10, if the SCTP packet rate is 250 packets/s, assume that each packet contains only a single data chunk and that the retransmission rate is 2% (the retransmission rate of the Internet is up to 5%), then achieved steganographic bandwidth is 10 bits/s.

In a similar way, multi-homing can be used for covert communication in the reverse direction (from User 2 to User 1). To accomplish this goal, SACK chunks with duplicate TSNs should be used.

## V. DETECTION POSSIBILITIES

For each of the groups of steganographic methods proposed in Section 3, detection or elimination solutions are sketched. Most of the proposed methods can be detected by comparing suspicious traffic to SCTP specifications or to statistics of non-steganographic SCTP traffic, as shown in Fig. 12. The main aim of this section is to point out potential enhancements that may be applied to SCTP standards to alleviate steganography utilisation, ideally at the standard development stage. Therefore, proposed countermeasures should be treated as guidelines for standard improvements.

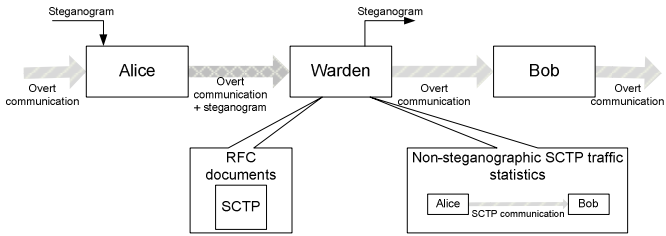

**Figure 12. Detection and elimination of steganogram**

### A. Methods that modify the content of SCTP packets

For steganographic methods that utilise modification to the SCTP packet content, possible detection techniques and proposed countermeasures are depicted in Table II.

TABLE II. POSSIBLE STANDARD IMPROVEMENTS TO NEUTRALISE STEGANOGRAPHIC METHODS THAT MODUFY CONTENT OF SCTP PACKETS

| Steg. method | Detection technique | Countermeasure (proposed standard change) |
|---|---|---|
| I1 | Analysis of Verification Tags values. | – |
| I2 | Comparison between values of Maximum Inbound Streams sent by "normal" users (users who do not use steganography) and suspicious user. | Limit possible values of Maximum Inbound Streams, i.e. only powers of 2 may be allowed. |
| D1 | Comparison between values of Stream Sequence Number sent by "normal" users (users who do not use steganography) and suspicious user. | For unordered transmission, Stream Sequence Number must be set to 0. |
| D2 | Checking value of Payload Stream Identifier. | Only standardised values must be allowed. |
| S1 | Analysis of a_rwnd values and sizes of received chunks. | – |
| S2 | Analysis of average number of duplicated chunks. | – |
| A1 | Analysis of Shared Key Identifier values. | Limit the number of shared keys for association to 1 or set one pair of shared keys for time slot, i.e. 10 minutes. |
| P1 | Analysis of Padding Data. | All bits of Padding Data must be set to 0. |
| VP1 | Checking the existence of IP addresses that are sent in these parameters. | Remove these parameters. Replace them with new chunk type, which will be sent from each user's address in order to add it to association. |
| VP2 | Comparison between values of Heartbeat Info Parameter sent by normal user (user who do not use steganography) and suspicious user. | Define value of Heartbeat Info Parameter. |
| VP3 | Analysis of RandomNumber. | – |
| VP4 | Comparison between values of ASCONF-Request Correlation ID sent by normal user (user who do not use steganography) and suspicious user. | ASCONF-Request Correlation ID must be a sequence number. |
| VP5 | Analysis of Padding Data. | All bits of Padding Data must be set to 0. |

### B. Methods that modify how SCTP packets are exchanged

#### B.1 MULTI-STREAMING

Similar to the multi-homing based steganographic method, detection of multi-streaming may be hard to perform and depends on the concrete application with which the SCTP will be utilised. If the pattern of stream usage is established, then statistical SCTP traffic analysis may reveal hidden communication.

Elimination of the proposed steganographic method may be achieved by changing TSNs by an intermediate node, e.g., an edge router with steganography detection functionality. Such operation may successfully interrupt the proper exchange of hidden data.

#### B.2 CHUNK-BASED FORMAT OF PACKETS

Detection of the first-described method based on influencing the chunk-based format of the packet may be hard to perform if constraints listed in the previous Section are satisfied. Otherwise, it will be quite easy to notice improper behaviour of the protocol. Elimination of such a method may be achieved by changing the order of chunks in the transmitted packets.

Detection of the second-described method, which uses a modular structure for the SCTP packet, may be hard to perform and depends on the concrete application for which the SCTP protocol will be utilised. Statistical analysis of the number of chunks in the SCTP packets can be helpful. Elimination of this method is possible by intentionally dividing each packet with more than one chunk into packets with smaller numbers of chunks, but this may have negative effects on packet delay variation.

### C. Hybrid method

If the number of intentionally omitted chunks is kept to a reasonable level, then detection of both proposed methods is difficult, and statistical analysis of the frequency of moving acknowledged TSNs may be helpful.

Elimination of the first method is possible by a specialised intermediate node that will be responsible for detection and the dropping of chunks that have already been acknowledged by the receiver.

Elimination of the second method is harder because the steganogram is sent before the FT chunk.

### D. Inter-protocol SCTP-based method

It is worth noting that steganographic methods that utilise multi-homing are generally harder to detect than single-homing ones. This is because, detecting covert communications requires observing the traffic on several different communication paths.

The resistance to the detection of methods proposed in Section 3 depends on how future typical SCTP implementations will behave. If alternative paths for retransmitted chunks often change, then the proposed steganographic method that utilises multi-homing will be harder to detect. However, if retransmitted chunks are sent through only one alternative path, then any other behaviour will be treated as an anomaly. Thus, the requirement that retransmitted chunks should be sent through only one alternative path should be enclosed in the SCTP standard.

Whatever the implementation, statistical analysis of NIC addresses used for retransmitted chunks may help to detect hidden communications. Elimination of proposed steganographic methods is possible by changing source and destination addresses of randomly chosen packets that contain retransmitted chunks.

## VI. IMPLEMENTATION

We decided to implement multi-streaming based steganographic methods to present practical aspects to SCTP-based steganography. This method was chosen because of the following reasons:
- The method has high steganographic bandwidth (one of the highest among the methods presented in this paper),
- The method is hard to detect (if it is used in appropriate applications),
- The method is potentially hard to eliminate.

Moreover, this method is easy to apply when the Linux operating system is used, and it can be implemented using the socket API for SCTP (high level programming) [13, 15], which provides access to new features of SCTP (including multi-streaming). However, other operating systems may require more sophisticated solutions. If an operating system is not using an FCFS scheduler, then the TSN numbers for chunks may not be assigned in the sequence of the *send()* function calls. For these cases, implementations may need to be done at a lower level.

The aim of the implemented MSD (Multi-streaming Steganographic Downloader) application is to simultaneously download several files from the server, and each file is downloaded within a different stream.

The most important part of the MSD is the multi-streaming based steganographic module. This module allows utilisation of all available SCTP streams, and allows for the setting of the size of the chunk groups, which are used to encode and decode hidden bits. The process of sending steganograms can be divided into three phases:
- Issuing the starting sequence.
- Sending the size of the used chunk groups.
- Sending the length of the steganogram and the steganogram itself.

### A. Phase 1: Issuing the starting sequence

The beginning of sending the steganogram is indicated by the starting sequence (the sequence of chunks from all streams in decreasing order of the streams' identifiers). For example, if five streams are used (five files are downloaded), and the identifiers of the streams are 0, 1, 2, 3, and 4, then the start sequence is five chunks, sent one by one, within streams 4, 3, 2, 1 and 0. The starting sequence is not sent before the steganogram is transmitted.

### B. Phase 2: Sending the size of the chunk groups

After the starting sequence, the size of the chunk groups, which are used to encode and decode the hidden bits, is sent. This size is defined by four subsequent chunks (an even identifier of the stream within which a chunk is sent means binary "0", and an odd identifier means binary "1").

### C. Phase 3: Sending the length of the steganogram and the steganogram

After establishment of the size of the chunk groups, the groups are used to code the length of the steganogram and the steganogram to be transmitted. The procedure of choosing subsequent groups of chunks that are to be sent is as follows:
1. All possible groups of chunks are sorted in increasing order.
2. Binary sequences of $\lfloor log_2(n^k) \rfloor$ elements ($n$ denotes the number of streams, and $k$ denotes the size of the chunk groups) are consecutively assigned to sorted chunk groups in lexicographical order.
3. If there are chunk groups without an assigned sequence, then sequences of $\lfloor log_2(n^k) \rfloor + 1$ elements, starting from sequence of zeros, are assigned to them.

## VII. EXPERIMENTAL METHODOLOGY

The implementations described for the multi-streaming based method were used to carry out an experiment. The objective was to choose an optimal size for the chunk groups, to convey the steganogram in the shortest time (using as few chunks as possible).

The number of chunks that must be sent to deliver a steganogram using a fixed number of streams may be different for various sizes of chunk groups used to code and decode hidden bits. Although the maximal bandwidth of this method does not depend on the size of the chunk groups, the lower bound of the bandwidth is different for various groups. The greater the lower bound of the bandwidth is, the greater is the guaranteed steganographic bandwidth. This relationship leads to the premise that groups for which the lower bound of the bandwidth is the greatest (Eq. (3-2)) will give the best results. The truth of this premise was checked in an experiment.

One of the most important aspects of this experiment is to choose a representative group of messages that can be sent as a steganogram. We decided that such a group is a list of words SOWPODS that is used in most Scrabble tournaments [16]. The list contains 267,751 words with lengths between two and fifteen letters [17]. Words were written with capital letters using the char type variable in the C programming language.

In the experiment, two cases were examined. In the former, *untapped bits* (bits that are not bits of a steganogram but that are sent within the last group of chunks that conveys a hidden message) are included, and, in the latter, untapped bits are not included. Examples of both cases is presented in Fig. 13.

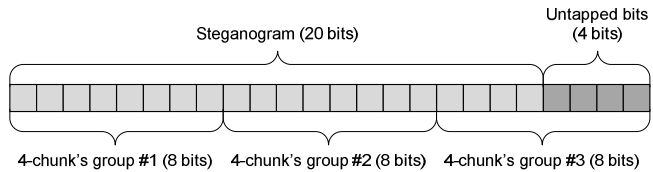

**Figure 13. Untapped bits example**

The length of the steganogram is 20 bits. Groups containing 4 chunks and 8 bits can be hidden in each group (in a case when 4 streams are used). Then, the last group (group #3) contains 4 bits of steganogram and 4 untapped bits. If untapped bits are included, then the number of the chunks needed to convey the steganogram is equal to the product of the sent groups and the length of a single group (in the example in Fig. 13 the result is 3·4=12 chunks). If untapped bits are not included, then the quotient of untapped bits and the lower bound of the steganographic bandwidth (Eq. (3-2)) is subtracted from the sent chunks (in the example in Fig. 13, the result is $3 \cdot 4 - 4 \div (\lfloor \log_2 4^4 \rfloor / 4) = 12 - 2 = 10$ chunks). Note that this case is an artificial case because untapped bits are indeed transmitted. The only objective of introducing this case is to help to draw correct conclusions.

In Fig. 14, a process of computation of measure used to show experimental results was presented. It is the number of words sent using a minimal number of chunks from among all transmitted words for the specified size of group and the specified number of streams.

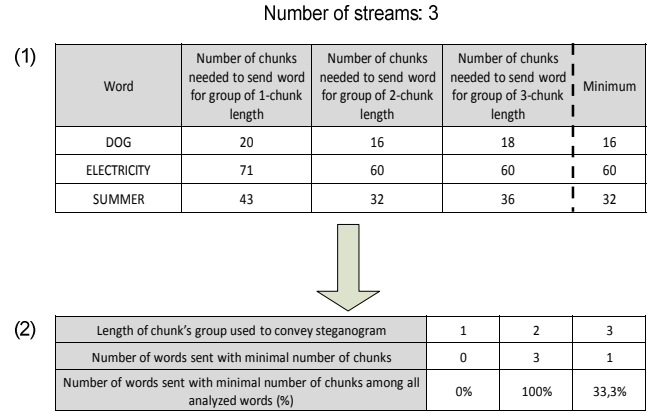

**Figure 14. Process of calculation of measure used to express experimental results**

In the example presented in Fig. 14, three streams are used. This example shows calculation of the measure for three words ("DOG", "ELECTRICITY", and "SUMMER") and three sizes of groups (1, 2, and 3 chunks). In the first step (1), the method checks how many chunks must be transmitted to convey each word as a steganogram, considering the size of each group. Moreover, the minimum number of chunks needed to send each word is computed, considering the sizes of the groups that we are concerned with. In the second step (2), the number of words sent using the minimal number of chunks is calculated for each size of chunk group. This value (expressed as a percentage) is used to compare experimental results.

The experiment was conducted for a number of streams between 2 and 15 and for sizes of chunk groups between 1 and 10. There was no need to consider more sizes of groups because steganograms used for experiments are short (the longest word has 15 letters, which equals to 120 bits).

## VIII. EXPERIMENTAL RESULTS

The results achieved in experiments are presented in Fig. 15-17. There are results for three chosen numbers of streams (2, 5 and 9). These cases allow formulating complete conclusions.

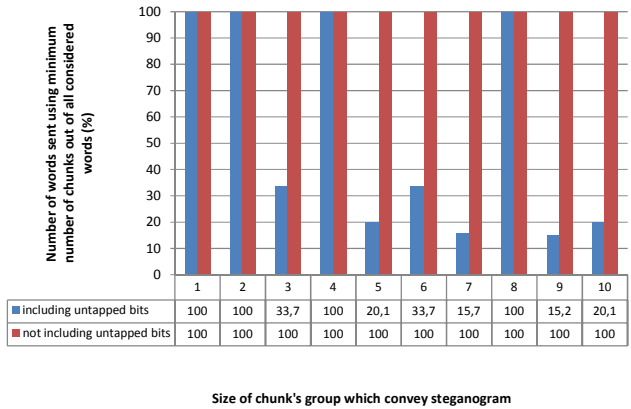

**Figure 15. Experimental result achieved for 2 streams**

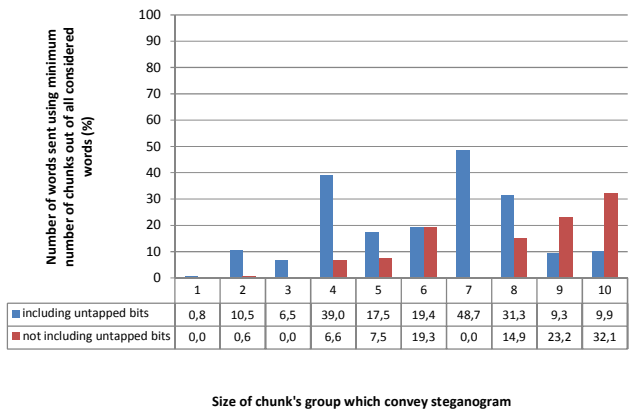

**Figure 16. Experimental result achieved for 5 streams**

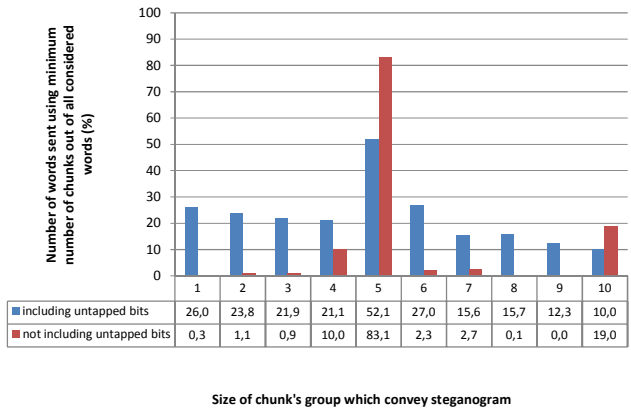

**Figure 17. Experimental result achieved for 9 streams**

The results for the case when 2 streams are used show that regardless of the size of the group, the same number of chunks is needed to send each steganogram if untapped bits are not included (Fig. 15). This result arises because the lower bound of the bandwidth and the maximum bandwidth are equal (Eqs. (3-1) and (3-2)). If untapped bits are included, then some steganograms are sent using more chunks than others. There is no difference between including and not including untapped bits only for chunk groups of sizes of 1, 2, 4 and 8. This similarity is caused by the fact that the length of each word used in the experiment is a multiple of 8 bits (each letter is written using 8 bits) and, in single groups of sizes of 1, 2, 4, 8 chunks, 1, 2, 4 and 8 hidden bits are transmitted, respectively. Results presented for the case when 2 streams are used (Fig. 15) are parallel to results for cases when the number of streams is any power of 2.

If the number of streams is a power of 2 (in the form of $2^p$, where $p$ is a natural number) and the steganogram is hidden using groups of size of $k$, then $\log_2(2^p)^k = k \cdot p$ bits are hidden in a single group. To send as few chunks as possible to transmit a steganogram, a chunk group size of 1 should be used. If the size of the group is greater, then the number of chunks that must be sent to convey the steganogram may be greater.

When 5 streams are used (Fig. 16) and untapped bits are not included, the best result is achieved for groups with a size of 10 (32.1%). This result is mainly caused by the fact that the lower bound of the bandwidth for this group is the greatest of all examined groups (2.3 bits/chunk). If untapped bits are included, the best result is achieved for a group of size of 7 (48.7%). The lower bound of the bandwidth is also high for this group (2.28 bits/chunk). This fact arises because of the many untapped bits in the case of a group size of 10. The lower bound of the bandwidth for this group is 2.3 bits/chunk; so, in a single group, 23 hidden bits are transmitted, and the lengths of the words are multiples of 8.

When 9 streams are used (Fig. 17), the best results for both cases are achieved for group with 5 chunks (52.1% when untapped bits are included, 83.1% when untapped bits are not included). It is worth noting that all groups of size greater than 5 have a greater lower bound of bandwidth. This result is caused mainly by two factors. The first factor is that there is a bigger influence of untapped bits on groups with a size greater than 5. The second factor is the increased probability of sending more bits than those that arise from the lower bound of bandwidth for groups with a smaller size (i.e., it is easier to fit a shorter sequence of bits among the bits of a steganogram).

The results for cases when 5 and 9 streams are used show that it is worthwhile to calculate the optimal size of a chunk group that is used to convey each steganogram, to send as few chunks as possible. The choice of the group for which the lower bound of the bandwidth is the greatest may not get the best result. Furthermore, the experimental results for cases with 5 and 9 streams show that only about 50% of the words are sent using a minimal number of chunks for the best group in the experiment. However, it is necessary to realise that analogous experiments for steganograms that are not English words can give different results.

The experimental results achieved show that only for cases when the number of streams is a power of 2 can the optimal size of chunk group required to send a steganogram, using as few chunks as possible be defined. For other numbers of streams, it is necessary to calculate the optimal group size for each steganogram.

## IX. CONCLUSIONS

In this paper, we proposed nineteen different steganographic methods that can be used in the SCTP protocol. The analysis presented in this paper takes into account all information about hiding possibilities for SCTP and points out possible countermeasures. All of these methods may lead to confidential information leakage and should be treated as a threat to network security. Many of them may be evaded by changing the SCTP standard – where it is possible, certain improvements were proposed.

For SCTP multi-streaming based steganographic methods, an experiment was performed using a developed application named MSD (Multi-streaming Steganographic Downloader). The experiment's main objective was to choose the optimal size for chunk groups to convey steganograms (English words) in the shortest time (using as few chunks as possible). The experimental results achieved proved that only for cases when the number of streams is a power of 2 can the optimal size of a chunk group to send a steganogram, using as few chunks as possible, be defined. For other numbers of streams, it is worthwhile to calculate the optimal group size for each steganogram.

Presented in this paper is a complete analysis on information hiding in the SCTP, and, as a result, this paper can be treated as a "guide" when developing steganalysis (detection) tools. This paper also emphasises how important it is to further inspect other network protocols that are to be utilised in future networks, to avoid hidden communications as early as possible, ideally, still at the standards development stage.

## ACKNOWLEDGMENTS

- The authors would like to thank the editors and the anonymous reviewers for their careful review of this paper and for their constructive remarks and suggestions.
- This work was partially supported by the Polish Ministry of Science and Higher Education under Grants: N517 071637 and IP2010 025470.